\documentclass[twocolumn,nofootinbib,amsmath,amssymb,a4paper]{revtex4}

\usepackage{graphicx}
\usepackage{dcolumn}
\usepackage{bm}

\newcommand{\bei}{\begin{itemize}}
\newcommand{\eei}{\end{itemize}}
\newcommand{\beq}{\begin{equation}}
\newcommand{\eeq}{\end{equation}}
\newcommand{\beqn}{\begin{eqnarray}}
\newcommand{\eeqn}{\end{eqnarray}}
\newcommand{\beqns}{\begin{eqnarray*}}
\newcommand{\eeqns}{\end{eqnarray*}}

\newcommand{\e}{\epsilon}
\def\bk{\!\!\!\!}

\def\ea{{\em et al.}}
\def\min{{\rm min}}

\def\rPTbarkappa{\kern 0.18em\overline{\kern -0.18em r}{}^{\kappa}{}}
\def\rPTbarsigma{\kern 0.18em\overline{\kern -0.18em r}{}^{\sigma}{}}

\def\deltabarkappa{\kern 0.18em\overline{\kern -0.18em \delta}{}_r^{\kappa}}
\def\deltabarsigma{\kern 0.18em\overline{\kern -0.18em \delta}{}_r^{\sigma}}
\def\deltaTbarkappa{\kern 0.18em\overline{\kern -0.18em \delta}{}_T^{\kappa}}
\def\deltaTbarsigma{\kern 0.18em\overline{\kern -0.18em \delta}{}_T^{\sigma}}

\newcommand\ph{\phantom}
\newcommand{\half}{\ensuremath{{1\over2}}}
\newcommand{\pvec}{{\bf p}}

\def\OC{X}
\def\OCbar{{\kern 0.18em\overline{\kern -0.18em \OC}}}
\def\mBz{m_{\Bz}}
\def\spz{s_{+}}
\def\smz{s_{-}}

\def\mpm{m_{0}}

\def\fpz{f_{+}}
\def\fmz{f_{-}}
\def\fpm{f_{0}}

\def\mpmMax{\mpm^{\rm max}}
\def\mpmMin{\mpm^{\rm min}}

\def\pipipi{\pip\pim\piz}
\def\Btopipipi{\Bz\to\pipipi}

\def\mprime{m^\prime}
\def\thetaprime{\theta^\prime}
\def\deprime{{\de^\prime}{}}
\def\dt{\deltat}

\def\de{\DeltaE}
\def\demin{\de_{-}}
\def\demax{\de_{+}}
\def\deminmax{\de_{\pm}}

\def\tpi{3\pi}

\def\detJ{|\det J|}

\def\a{\kappa}

\def\Amptp{{A}_{3\pi}}
\def\Amptpbar{\kern 0.18em\overline{\kern -0.18em {\cal A}}_{3\pi}}

\def\absAmptp{|\Amptp|}
\def\absAmptpbar{|\Amptpbar|}
\def\Amptpkappa{{A^{\kappa}}}
\def\Amptpsigma{{A^{\sigma}}}
\def\Amptpbarkappa{\kern 0.18em\overline{\kern -0.18em A}{}^{\kappa}{}}
\def\Amptpbarsigma{\kern 0.18em\overline{\kern -0.18em A}{}^{\sigma}{}}
\def\AmpAll{|{\cal A}_{3\pi}^\pm(\dmt)|^2}

\def\Tbarkappa{\kern 0.18em\overline{\kern -0.18em T}{}^{\kappa}{}}
\def\Tbarsigma{\kern 0.18em\overline{\kern -0.18em T}{}^{\sigma}{}}
\def\Pbarkappa{\kern 0.18em\overline{\kern -0.18em P}{}^{\kappa}{}}
\def\Pbarsigma{\kern 0.18em\overline{\kern -0.18em P}{}^{\sigma}{}}
\def\kappab{\overline\kappa}

\def\ji{\kappab}
\def\ij{\kappa}
\def\Aij{{A^{\ij}}}
\def\Abij{\kern 0.18em\overline{\kern -0.18em A}{}^{\ij}{}}
\def\Tij{{T^{\ij}}}
\def\Tji{{T^{\ji}}}
\def\Pij{{P^{\ij}}}
\def\Pji{{P^{\ji}}}
\def\R{{\rm Re}}
\def\I{{\rm Im}}
\def\kappm{\kappa^{+-}}
\def\kapmp{\kappa^{-+}}

\def\bbar{\Bz \Bzb}

\def\Crhopi{C}
\def\dCrhopi{\Delta C}

\def\dS{\Delta S}
\def\dC{\Delta C}

\def\Acp{{\cal A}_{\rho\pi}}

\def\Acppm{{\cal A}_{\rho\pi}^{+-}}
\def\Acpmp{{\cal A}_{\rho\pi}^{-+}}

\def\Nbpm{{\kern 0.18em\overline{\kern -0.18em N}}^{+-}}
\def\Nbmp{{\kern 0.18em\overline{\kern -0.18em N}}^{-+}}
\def\rar{\rightarrow}

\def\Mu{\mu}
\def\Chi2MinaMu{\chi^2_{\min ;\a,\Mu}}
\def\Chi2MinMu{\chi^2_{\min ;\Mu}(a)}

\def\dmt{\Delta t}
\def\dmd{\Delta m_d}

\def\fscfave{\kern 0.18em\overline{\kern -0.18em f}_{\rm SCF}}
\def\fscf{f_{\rm SCF}}

\def\abar{\bar{a}}

\def\Bbar{\kern 0.18em\overline{\kern -0.18em B}{}\xspace}

\def\BRpmb{{\cal \kern 0.18em\overline{\kern -0.18em  B}}{}_{\rho\pi}^{+-}}
\def\BRmpb{{\cal \kern 0.18em\overline{\kern -0.18em  B}}{}_{\rho\pi}^{-+}}

\def\BRipmb{{\cal \kern 0.18em\overline{\kern -0.18em  B}}{}_{\rho^+\pi^-}}
\def\BRimpb{{\cal \kern 0.18em\overline{\kern -0.18em  B}}{}_{\rho^-\pi^+}}

\def\Abar{\kern 0.18em\overline{\kern -0.18em A}{}}
\def\abar{\kern 0.18em\overline{\kern -0.18em a}{}}

\def\Apm{A^{+}}
\def\Amp{A^{-}}
\def\Apmb{\Abar^{+}}
\def\Ampb{\Abar^{-}}

\def\Azz{A^{0}}
\def\Azzb{\Abar^{0}}

\def\ie{{\em i.e.}} 
\def\cf{{\em cf.}}

\input{babarsym}

\begin{document}

\title{ $\Bz\to\pip\pim\piz$ Time Dependent Dalitz analysis at BaBar.}

\author{Gianluca Cavoto}
\email{gianluca.cavoto@roma1.infn.it}
\affiliation{INFN Sezione di Roma, Piazzale Aldo Moro 2, 00185 Rome, Italy }%

\begin{abstract}
  I present here results of a time-dependent analysis of the Dalitz structure of neutral $B$ meson decays to 
 $\pip\pim\piz$  from a dataset of  346 million $B \bar B$ pairs collected at the $\Upsilon(4S)$ center of mass energy by the BaBar detector at the SLAC PEP-II $e^+e^-$ accelerator.  No significant CP violation  effects are  observed and  68\% confidence interval is derived on the weak angle $\alpha$  to be [75$^¡$,152$^¡$] 
\end{abstract}

\maketitle

\section{Introduction}
\label{sec:introduction}
The   time-dependent   analysis of
 the $\Bz\to\pip\pim\piz$ Dalitz plot (DP),   dominated 
by the $\rho(770)$ intermediate resonances,
 extracts simultaneously   the strong transition 
amplitudes and the weak interaction phase 
$\alpha\equiv \arg\left[-V_{td}^{}V_{tb}^{*}/V_{ud}^{}V_{ub}^{*}\right]$
of the Unitarity Triangle ~\cite{SnyderQuinn}. In the Standard Model, a non-zero 
value for $\alpha$ is responsible for the occurrence 
of mixing-induced \CP violation in this decay.
$\rho^{\pm}\pi^{\mp}$ is not a \CP 
eigenstate, and four flavor-charge configurations
$(\Bz(\Bzb) \to \rho^{\pm}\pi^{\mp})$ must be considered.  
The corresponding isospin analysis~\cite{Lipkinetal} is unfruitful
with the present statistics since two pentagonal amplitude relations with 
12 unknowns have to be solved (compared to 6 unknowns for
the  $\pip\pim$ and $\rho^+\rho^-$ systems).

The differential $\Bz$ decay width with respect to the 
Mandelstam variables $\spz$, $\smz$ (\ie, the 
{\em Dalitz plot} \cite{pdg2006}) reads $	d\Gamma(\Btopipipi)  = \frac{1}{(2\pi)^3}\frac{|\Amptp|^2}{8 \mBz^3}\,d\spz d\smz$, 
where $\Amptp$  ($\Amptpbar$) is the Lorentz-invariant amplitude
of the three-body decay  $\Bz\to\pip\pim\piz$ ($\Bzb\to\pip\pim\piz$). 
We assume in the following that the amplitudes  are dominated by the three 
resonances $\rho^+$, $\rho^-$ and $\rho^0$ and we write
$   \Amptp    	= \fpz \Apm + \fmz \Amp + \fpm\Azz $  and 
$   \Amptpbar 	 = \fpz \Apmb + \fmz \Ampb + \fpm\Azzb $,
 where the $f_\kappa$ (with $\kappa=\{+,-,0\}$ denoting the charge of the
$\rho$ from the decay of the $\Bz$ meson) 
are functions of 
$\spz$ and $\smz$ that incorporate the kinematic and dynamical properties 
of the $\Bz$ decay into a (vector) $\rho$ resonance and a 
(pseudoscalar) pion, and where the $\Aij$ are 
complex amplitudes
that include weak and strong transition phases and that are independent 
of the Dalitz variables.

With $\deltat \equiv t_{\tpi} - t_{\rm tag}$ defined as the proper 
time interval between the decay of the fully reconstructed $B^0_{\tpi}$ 
and that of the  other meson $\Bz_{\rm tag}$,  the time-dependent decay
rate  when the tagging meson is a $\Bz$ ($\Bzb$) 
is given by 
\beqn
\label{eq:dt}
    \AmpAll
	=
    \frac{e^{-|\dmt|/\tau_{B^0}}}{4\tau_{B^0}}  \nonumber \\
	\bigg[\absAmptp^2 + \absAmptpbar^2
	      \mp \left(\absAmptp^2 - \absAmptpbar^2\right)\cos(\dmd\dmt)
	      \nonumber \\
	      \pm\,2\I\left[\Amptpbar\Amptp^*\right]\sin(\dmd\dmt)	
	\bigg]~,
\eeqn
where $\tau_{B^0}$ is the mean \Bz lifetime and $\deltamd$ is the $\BzBzb$ 
oscillation frequency. Here, we have assumed that \CP violation in $\bbar$ 
mixing is absent ($|q/p|=1$), $\Delta\Gamma_{B_d}=0$ and \CPT is conserved. 
 Inserting the amplitudes  $\Amptp   $  and $\Amptpbar$ one 
obtains for the terms in Eq.~(\ref{eq:dt})
\beqn
   \label{eq:UI}
   \absAmptp^2 \pm \absAmptpbar^2 
	=
	\sum_{\kappa\in\{+,-,0\}}  |f_\kappa|^2U_\kappa^\pm
	\;\;+ \;\;  \nonumber \\
	2\bk\bk\sum_{\kappa <\sigma\in\{+,-,0\}} 
	\left(
	    \,\R\left[f_\kappa f_\sigma^*\right]U_{\kappa\sigma}^{\pm,\R}
	  - \,\I\left[f_\kappa f_\sigma^*\right]U_{\kappa\sigma}^{\pm,\I}
	\right)~,
	\nonumber\\[0.2cm]
   \I\left(\Amptpbar\Amptp^*\right)
	=
	\sum_{\kappa\in\{+,-,0\}}  |f_\kappa|^2I_\kappa
	\;\;+  \nonumber \\
	\sum_{\kappa <\sigma\in\{+,-,0\}} 
	\left(
	    \,\R\left[f_\kappa f_\sigma^*\right]I_{\kappa\sigma}^{\I}
	  + \,\I\left[f_\kappa f_\sigma^*\right]I_{\kappa\sigma}^{\R}
	\right)~,	
\eeqn

The 27 real-valued  coefficients  defined in Tab.\ref{tab:resultsUandI}
that multiply the $f_\kappa f_\sigma^*$ bilinears  are determined by the fit. Each of the coefficients 
is related in a unique way to physically more intuitive quantities,
such as tree-level and penguin-type amplitudes, the angle $\alpha$, or
the quasi-two-body \CP and dilution parameters~\cite{rhopipaper} 
(\cf\   Section~\ref{subsec:interpretation}).  
We determine the quantities of interest in a subsequent least-squares
fit to the measured $U$ and $I$ coefficients.

\section{Dalitz Model}
\label{sec:dalitzmodel}
The $\rho$ resonances
are assumed to be the sum of the ground state $\rho(770)$ and the
radial excitations $\rho(1450)$ and $\rho(1700)$, with 
resonance parameters determined by a combined fit 
to $\tau^+\to\nutb\pip\piz$ and $\epem\to\pip\pim$ data~\cite{taueeref}.
Since the hadronic environment is different in \B decays, we 
cannot rely on this result and therefore determine the relative $\rho(1450)$
and $\rho(1700)$ amplitudes simultaneously with the \CP parameters from the fit. Variations of
the other parameters and possible contributions to the $\Bz\to\pip\pim\piz$ 
decay other than the $\rho$'s are studied as part of the systematic 
uncertainties (Section~\ref{sec:Systematics}).

Following Ref.~\cite{taueeref}, the $\rho$
resonances are parameterized in  $f_\kappa$ by a modified relativistic Breit-Wigner 
function introduced by Gounaris and Sakurai (GS)~\cite{rhoGS}. 

Large variations occurring in small areas of the Dalitz plot are very difficult to describe in detail.  
These regions are particularly important since this is where the interference, and hence our ability to determine the strong phases, occurs. 
We therefore apply the transformation $	d\spz \,d\smz \;\longrightarrow \detJ\, d\mprime\, d\thetaprime$,
which defines the {\em Square Dalitz plot} (SDP). The new coordinates 
are $\mprime \equiv \frac{1}{\pi} \arccos\left(2\frac{\mpm - \mpmMin}{\mpmMax - \mpmMin} - 1 \right),~\thetaprime \equiv \frac{1}{\pi}\theta_{0}$, 
where $\mpm$ is the invariant mass between the charged tracks,
$\mpmMax=\mBz - m_{\pi^0}$ and $\mpmMin=2m_{\pi^+}$ are the kinematic
limits of $\mpm$ and $\theta_{0}$ is the $\rho^0$ helicity angle;
$\theta_{0}$ is defined by the angle between the $\pi^+$ in 
the $\rho^0$ rest frame and the $\rho^0$ flight direction in 
the $\Bz$ rest frame. 
\begin{figure}
\includegraphics[width=0.4\textwidth]{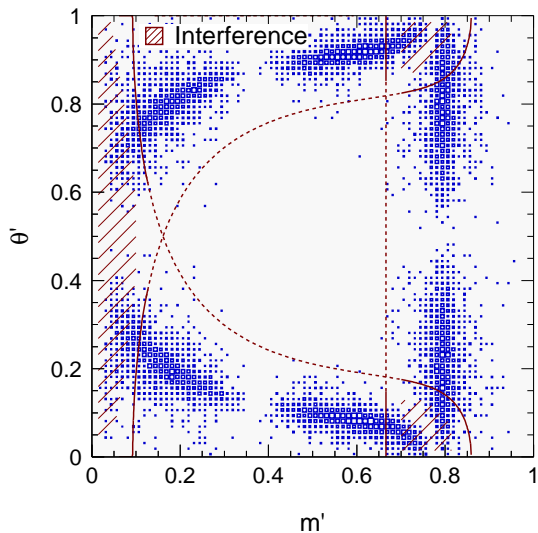}
\caption{\label{fig:squareDalitz} Square   Dalitz plots for Monte-Carlo
	generated $\Bz\rar\pi^+\pi^-\pi^0$ decays.The decays have been
	simulated without any detector effect and the amplitudes $\Apm$,
	$\Amp$ and $\Azz$ have all been chosen equal to 1 in order to
	have destructive interferences at equal $\rho$ masses. The main
	overlap regions between the charged and neutral $\rho$ bands
	are indicated by the hatched areas.
	Dashed lines in both plots correspond to
	$\sqrt{s_{+,-,0}}=1.5~{\rm GeV}/c^2$: the central region of the Dalitz
	plot  contains almost no signal event.
}
\end{figure}
$J$ is the Jacobian of the transformation 
that zooms into the kinematic boundaries of the Dalitz plot, shown in Fig.\ref{fig:squareDalitz} .


\section{Analysis description}
\label{sec:analysisdescription}

The $U$ and $I$ coefficients and the $\Btopipipi$ event yield are
determined by a maximum-likelihood fit of the signal model to the 
selected candidate events. Kinematic and event shape variables 
exploiting the characteristic properties of the events are used 
in the fit to discriminate signal from background. 

\subsection{Signal and background parametrization}

We reconstruct $\Btopipipi$ candidates from pairs of 
oppositely-charged tracks, which are required to form a good quality vertex,
and a $\pi^0$ candidate.  In order to ensure that all events are within 
the Dalitz plot boundaries, we constrain the three-pion invariant mass to the $B$ mass.

A $B$-meson candidate is characterized kinematically by the energy-substituted 
mass $\mes=\lbrack{(\half s+\pvec_0\cdot\pvec_B)^2/E_0^2-\pvec_B^2}\rbrack^\half$
and energy difference $\de = E_B^*-\half\sqrt{s}$, 
where $(E_B,\pvec_B)$ and $(E_0,\pvec_0)$ are the four-vectors
of the $B$-candidate and the initial electron-positron system,
respectively. The asterisk denotes the \FourS\  frame,
and $s$ is the square of the invariant mass of the electron-positron system.  
We require $5.272 < \mes <5.288\gevcc$. 
The $\de$ resolution 
exhibits a dependence on the $\pi^0$ energy and therefore varies 
across the Dalitz plot. We account for this effect by introducing
the transformed quantity $\deprime=(2\de - \demax - \demin)/(\demax - \demin)$,
with $\deminmax(\mpm)=c_{\pm}-\left(c_{\pm}\mp\bar c\right)(\mpm/\mpmMax)^2$,
where $\mpm$ is strongly correlated with the energy of $\piz$. 
We use the values
$\bar c = 0.045\gev$, $c_{-} = -0.140\gev$, $c_{+} = 0.080\gev$,
$\mpmMax = 5.0\gev$, and require $-1<\deprime<1$. 

Backgrounds arise primarily from random combinations in continuum  $q \bar q$ events.
To enhance discrimination between signal and continuum, we 
use a neural network (NN)~\cite{NNo} to combine  discriminating topological  variables.

The time difference $\deltat$ is obtained from the measured distance between 
the $z$ positions (along the beam direction) of the $\Bz_{\tpi}$ and 
$\Bz_{\rm tag}$ decay vertices, and the boost $\beta\gamma=0.56$ of 
the \epem\ system: $\deltat = \Delta z/\beta\gamma c$.
To determine the flavor of the $\Bz_{\rm tag}$ 
we use the $B$ flavor tagging algorithm of Ref.~\cite{BabarS2b}.
This produces six mutually exclusive tagging categories.

Events with multiple \B 
candidates passing the full selection occur 
in $16\%$ $(\rho^\pm\pi^\mp)$ and $9\%$ $(\rho^0\pi^0)$ 
of the time, according to signal MC. 
If the multiple candidates have different $\pi^0$ candidates, 
we choose the B candidate with the reconstructed $\pi^0$ mass closest 
to the nominal $\pi^0$ mass; 
in the case that both candidates have the same $\pi^0$, we pick the first one.

The signal efficiency determined from MC simulation is $24\%$ for 
$B^0 \to \rho^\pm\pi^\mp$ and $B^0 \to \rho^0\pi^0$ events, and 
$11\%$ for non-resonant $\Btopipipi$ events. 

Of the selected signal events, $22\%$ of $B^0 \to \rho^\pm\pi^\mp$, 
$13\% $ of $B^0 \to \rho^0\pi^0$, and $6\%$ of non-resonant events are 
misreconstructed.  Misreconstructed events occur when a track or 
neutral cluster from the tagging $B$ is assigned to the reconstructed signal candidate. 
This occurs most often for  low-momentum tracks and photons and hence the misreconstructed events 
are concentrated in the corners of the Dalitz plot.  Since these are also the areas where the $\rho$ resonances
overlap strongly, it is important to model the misreconstruced events correctly.  

We use MC simulated events to study the background from other $B$ 
decays. More than a hundred channels were considered in 
preliminary studies, of which twenty-nine are  included
in the final likelihood model.  For each mode, the expected number of selected events is
computed by multiplying the selection efficiency (estimated using MC
simulated decays) by the world average branching fraction (or upper limit), scaled to the dataset 
luminosity ($310\;\mathrm{fb}^{-1}$). 
The selected on-resonance data sample is assumed to consist of signal, 
continuum-background and \B-background components, separated by the 
flavor and tagging category of the tag side \B decay. 
The signal likelihood consists of the sum of a correctly 
reconstructed (``truth-matched'', TM) component and a misreconstructed 
(``self-cross-feed'', SCF) component.

\subsection{Dalitz  and $\Delta t$ distribution}
\label{subsec:dalitzdistribution}

	The Dalitz plot PDFs require as input the Dalitz plot-dependent 
	relative selection efficiency, $\e=\e(\mprime,\thetaprime)$, 
	and SCF fraction, $\fscf=\fscf(\mprime,\thetaprime)$.
	Both quantities are taken from MC simulation. 
	
	Away from the Dalitz plot corners the efficiency is uniform, while it 
	decreases when approaching the corners, where one  of the 
	three particles in the final state is close to rest so that the 
	acceptance requirements on the particle reconstruction become 
        restrictive.
	Combinatorial backgrounds and hence SCF fractions are large in
	the corners
	of the Dalitz plot due to the presence of soft neutral clusters 
	and tracks. 
	
	The width of the dominant $\rho(770)$ resonance is large compared 
	to the mass resolution for TM events (about $8\mevcc$ core Gaussian
	resolution). We  therefore neglect resolution effects in the TM 
	model.	
	Misreconstructed events	have a poor mass resolution that strongly 
	varies across the Dalitz plot. It is described in the fit by a 
	$2\times 2$-dimensional resolution function, convoluted with signal Dalitz PDF.

	The $\deltat$ resolution function for signal and \B-background 
	events is a sum of three Gaussian distributions, with parameters 
	determined by a fit to fully reconstructed $\Bz$ 
	decays~\cite{BabarS2b}. 

	The Dalitz plot- and $\dt$-dependent PDFs factorize for the 
	charged-$B$-background modes, but not necessarily
	for the neutral-$B$ background due to $\BzBzb$ mixing.

 	The charged \B-background
		contribution to the likelihood parametrizes  tag-``charge'' 
		correlation (represented by an effective 
		flavor-tag-versus-Dalitz-coordinate correlation),
		and therefore possible direct \CP violation in these events.

		The  Dalitz plot PDFs are obtained from MC simulation and are 
		described with the use of non-parametric functions.
		The $\dt$ resolution parameters are determined by a fit to fully 
		reconstructed $\Bp$ decays.  

	       The neutral-$B$ background is parameterized with PDFs that
		depend on the flavor tag of the event and, depending on the final states they can show 
		 correlations between the flavor tag and the Dalitz 
		coordinate. 		
			The Dalitz plot PDFs are obtained from MC simulation and are 
		described with the use of non-parametric functions.
		For neutral-$B$ background, the signal $\dt$ resolution model 
		is assumed.

	The Dalitz plot
	     of the continuum events is parametrized with an empirical shape. 
		extracted  from on-resonance events selected in the
		$\mes$ sidebands and corrected for feed-through
		from \B decays.  
		The continuum $\deltat$ distribution is parameterized as the sum of 
		three Gaussian distributions with common mean and 
		three distinct widths that scale the $\dt$ per-event error, all determined by the fit.

\section{Results}
\label{sec:results}

The maximum-likelihood fit results in a $\Btopipipi$ event yield of
$1847\pm69$, where the error is statistical only. For the $U$ and $I$
coefficients, the results are given together with their statistical
and systematic errors in Table~\ref{tab:resultsUandI}. 
The signal is 
dominated by $\Bz\to\rho^\pm\pi^\mp$ decays. We observe an excess of
$\rho^0\piz$ events, which is in agreement with our previous upper
limit~\cite{BABARrho0pi0}, and the latest measurement from the Belle 
collaboration~\cite{BELLErho0pi0}.
The result for the $\rho(1450)$ amplitude is in agreement with the findings 
in $\tau$ and $\epem$ decays~\cite{taueeref}. For the relative strong phase between 
the $\rho(770)$ and the $\rho(1450)$ amplitudes we find 
$(171\pm23)^\circ$ (statistical error only), which is 
compatible with the result from $\tau$ and $\epem$ data.

\begin{table}[h]
\begin{center}
\begin{tabular}{cc} 
\hline
\hline
\multicolumn{2}{c}{ "Quasi twobody" $U_\kappa^\pm  = |\Amptpkappa|^2 \pm |\Amptpbarkappa|^2$} \\ \hline
$U_0^+$   $\rho^0 \pi^0$ fit fraction  &   $\phantom{-}0.237\pm0.053\pm0.043$ \\
$U_-^+$    $\rho^-\pi^+$ fit fraction   &   $\phantom{-}1.33\pm0.11\pm0.04$  \\ \hline
$U_0^-$  Direct CPV  ($\rho^0 \pi^0$) &     $-0.055\pm0.098\pm0.13$  \\
$U_-^-$  Direct CPV  ($\rho^- \pi^+$) &    $-0.30\pm0.15\pm0.03$  \\
$U_+^-$  Direct CPV  ($\rho^+ \pi^-$) &  $\phantom{-}0.53\pm0.15\pm0.04$  \\ 
\hline\hline
\multicolumn{2}{c}{ "Quasi twobody"  $I_\kappa = \I\left[\Amptpbarkappa\Amptpkappa{}^*\right]$} \\ \hline\hline 
$I_0$    Int. Mixing CPV  $\rho^0 \pi^0$ & $-0.028\pm0.058\pm0.02$ \\
$I_-$     Int. Mixing CPV  $\rho^- \pi^+$  & $-0.03\pm0.10\pm0.03$ \\
$I_+$    Int. Mixing CPV  $\rho^+ \pi^-$  &  $-0.039\pm0.097\pm0.02$ \\ 
\hline \hline
\multicolumn{2}{c}{ "Interference" $ U_{\kappa\sigma}^{\pm,\R(\I)}= \R(\I)\left[\Amptpkappa \Amptpsigma{}^*     \pm \Amptpbarkappa \Amptpbarsigma{}^*\right]$} \\  
\hline
$U_{+-}^{+,\I}$     & $\phantom{-}0.62\pm0.54\pm0.72$   \\
$U_{+-}^{-,\I}$  
    & $\phantom{-}0.13\pm0.94\pm0.17$  \\
$U_{+-}^{+,\R}$ 
     & $\phantom{-}0.38\pm0.55\pm0.28$    \\
$U_{+-}^{-,\R}$ 
    & $\phantom{-}2.14\pm0.91\pm0.33$   \\
$U_{+0}^{+,\I}$ 
  & $\phantom{-}0.03\pm0.42\pm0.12$   \\
$U_{+0}^{+,\R}$ 
    & $-0.75\pm0.40\pm0.15$    \\
$U_{+0}^{-,\I}$
     & $-0.93\pm0.68\pm0.08$  \\
$U_{+0}^{-,\R}$ 
     & $-0.47\pm0.80\ph{0}\pm0.3\ph{0}$ \\
$U_{-0}^{+,\I}$ 
     & $-0.03\pm0.40\pm0.23$ \\
$U_{-0}^{+,\R}$ 
     & $-0.52\pm0.32\pm0.08$    \\
$U_{-0}^{-,\I}$ 
     & $\phantom{-}0.24\pm0.61\ph{0}\pm0.2\ph{0}$  \\
$U_{-0}^{-,\R}$ 
 & $-0.42\pm0.73\pm0.28$   \\
 \hline\hline
     \multicolumn{2}{c}{  "Interference" $I_{\kappa\sigma}^{\R}  = \R\left[\Amptpbarkappa\Amptpsigma{}^*     - \Amptpbarsigma\Amptpkappa{}^*\right]$}  \\
       \hline 
    $I_{+-}^{\R}$ 
   & $-0.1\ph{0}\pm1.9\ph{0}\pm0.3\ph{0}$\\
$I_{+0}^{\R}$ 
     & $\phantom{-}0.2\ph{0}\pm1.1\ph{0}\pm0.4\ph{0}$ \\
 $I_{-0}^{\R}$ 
    & $\phantom{-}0.92\pm0.91\ph{0}\pm0.4\ph{0}$ \\
\hline \hline
\multicolumn{2}{c}{ "Interference"  $I_{\kappa\sigma}^{\I}   = \I\left[\Amptpbarkappa\Amptpsigma{}^*     + \Amptpbarsigma\Amptpkappa{}^*\right] $} \\ 
\hline
 $I_{+-}^{\I}$ 
    & $-1.9\ph{0}\pm1.1\ph{0}\pm0.1\ph{0}$  \\
$I_{+0}^{\I}$ 
     & $-0.1\ph{0}\pm1.1\ph{0}\pm0.3\ph{0}$  \\
     $I_{-0}^{\I}$ 
   & $\phantom{-}0.7\ph{0}\pm1.0\ph{0}\pm0.3\ph{0}$  \\
   \hline \hline
\end{tabular}
\label{tab:resultsUandI}
\end{center}
   \caption{ Definitions and results for the 26  $U$ and $I$  observables  extracted from the fit. We determine the relative values of $U$ and $I$ coefficients 
to $U_+^+$.}
\end{table}

\subsection{ Systematics studies}
\label{sec:Systematics}

The most important contribution to the systematic uncertainty stems
from the modeling of the Dalitz plot dynamics for signal. 
We evaluated this  by observing the difference
between the true values and Monte Carlo fit results, in which
events are generated based on an alternative model. 
 The alternative  fit model   has,  in addition, a uniform Dalitz distribution
for the non-resonance events and  possible resonances
including $f_0(980)$, $f_2(1270)$, and a low mass $S$-wave $\sigma$. 
 The fit does not find significant number of any of those decays.
However, the inclusion of a low mass $\pipi$ $S$-wave
component significantly degrades our ability to identify $\rho^0\piz$ events.
.

We vary the mass and width of the $\rho(770)$, $\rho(1450)$, and $\rho(1700)$ 
within ranges that 
exceed twice the errors found for these parameters in the fits to 
$\tau$ and $\epem$ data~\cite{taueeref}, and assign the observed
differences in the measured $U$ and $I$ coefficients as systematic uncertainties.

To validate the fitting tool, we perform fits on large MC samples with
the measured proportions of signal, continuum and $B$-background events.
No significant biases are observed in these fits, and the statistical
uncertainties on the fit parameters are taken as systematic uncertainties

Another major source of systematic uncertainty is the $B$-background model. 
The expected event yields from the background modes are varied according 
to the uncertainties in the measured or estimated branching fractions
Since $B$-background modes may exhibit  \CP violation, the corresponding 
parameters are varied within appropriate uncertainty ranges.

 Continuum  Dalitz 
plot PDF is extrapolated form \mes sideband, and  large 
samples of off-resonance data  with loosened  requirements on \de and 
the NN are used to   compare the distributions of $\mprime$ and $\thetaprime$ 
between the \mes sideband and the signal region. No significant 
differences are found. We assign as systematic error the effect seen when
weighting the continuum Dalitz plot PDF by the ratio of both data 
sets. This effect is mostly statistical in origin.

Other  systematic effects due to the signal PDFs 
comprise uncertainties in the
PDF parameterization, the treatment of misreconstructed events, the
tagging performance, and the modeling of the signal contributions and are estimated using arious data control samples.

\subsection{Intepretation of the results}
\label{subsec:interpretation}

\begin{figure}
\includegraphics[width=0.4\textwidth]{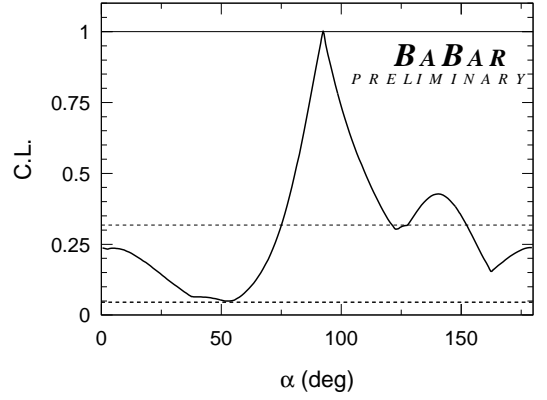}
  \caption{\label{fig:deltaalpha} 
	Confidence level functions for  
	$\alpha$. Indicated 
	by the dashed horizontal lines are the confidence level (C.L.) values
        corresponding to $1\sigma$ and $2\sigma$, respectively.}
\end{figure}

The $U$ and $I$ coefficients are related to the 
quasi-two-body parameters (Tab.\ref{tab:q2bresults}) defined in Ref.~\cite{rhopipaper},  explicitly accounting for the presence of interference effects, and are  thus exact even for a $\rho$ with finite width.
The systematic errors are dominated by the 
uncertainty on the \CP content of the \B-related backgrounds. 
One can transform the experimentally convenient, namely
uncorrelated, direct \CP-violation parameters $\Crhopi$ and $\Acp$
into the physically more intuitive quantities $\Acppm$ and $\Acpmp$.
The significance, including 
systematic uncertainties and calculated by using a mininum $\chi^2$
method, for the observation of non-zero direct \CP violation is 
at the $3.0\sigma$ level. 

\begin{table}[h]
\begin{center}
\begin{tabular}{cc} 
\hline
\hline
$C=(C^++C^-)/2$ &  $ \ph{-}0.154 \pm 0.090 \pm 0.037 $  \\
$S=(S^++S^-)/2$ & $\ph{-}0.01\pm 0.12\pm 0.028$  \\
 $\dC=(C^+-C^-)/2$ & $ 0.377\pm 0.091\pm 0.021$ \\
  $\dS=(S^+-S^-)/2$  & $0.06\pm 0.13\pm 0.029$ \\
$\Acp = \frac{ U^+_+ \, - U^+_- }{ U^+_+ \, + U^+_- }$  &  $-0.142\pm 0.041 \pm{0.015}$ \\ 
$  \Acppm =  \frac{|\kappm|^2-1}{|\kappm|^2+1}$ &  $0.03\pm0.07\pm0.03$  \\ 
$\Acpmp = \frac{|\kapmp|^2-1}{|\kapmp|^2+1}$ &  $ -0.38^{\,+0.15}_{\,-0.16}\pm0.07$ \\ 
\hline\hline 
\end{tabular} 
\label{tab:q2bresults}
\caption{ Quasi twobody parameters definition and results, where $C^{\pm}= \frac{ U^-_{\pm} }{ U^+_{\pm}}  $ and  $S^{\pm} = \frac{ 2 \, I_{\pm}}{ U^+_{\pm} }  $; $\kappm = (q/p)(\Ampb/\Apm)$ and $\kapmp = (q/p)(\Apmb/\Amp)$, so that $\Acppm$ ($\Acpmp$) involves only diagrams where the $\rho$ ($\pi$) meson is emitted by the $W$ boson. $\Acppm$ and $\Acpmp$  are evaluated as   $-\frac{\Acp+\Crhopi+\Acp\dCrhopi}{1+\dCrhopi+\Acp\Crhopi}$ and $\frac{\Acp-\Crhopi-\Acp\dCrhopi}{1-\dCrhopi-\Acp\Crhopi}$. Their  correlation coefficient is 0.62.  }
\end{center}
\end{table}

The measurement of the resonance interference terms allows us to  constrain the relative phase $\delta_{+-} = \arg\left( A^{+*}A^{-} \right)$ between the amplitudes of the  decays $B^0\to\rho^-\pi^+$ and $B^0\to\rho^+\pi^-$.  This constraint can be improved with the use of strong  isospin symmetry. The amplitudes  $\Aij$ represent the sum of tree-level $(\Tij)$  and penguin-type  $(\Pij)$
amplitudes, which have different CKM factors. Here we denote by $\kappab$ the charge conjugate of $\kappa$, where $\overline 0=0$.  We define ~\cite{BaBarPhysBook} $\Aij 		= \Tij e^{-i\alpha} + \Pij $ and $\Abij 		= \Tji e^{+i\alpha} + \Pji$,
where the magnitudes of the CKM factors have been absorbed in the 
$\Tij$, $\Pij$, $\Tji$ and $\Pji$.  Using
strong isospin symmetry and neglecting isospin-breaking effects,
one can identify $P^{0}=-(P^{+}+P^{-})/2$ and 9   unknowns have to be determined by the fit. 

We find for the solution that is favored by the fit $ \delta_{+-} \; = \; \left(34\,\pm29\right)^\circ$,
where the errors include both statistical and systematic effects, but only a marginal constraint on $\delta_{+-}$ is obtained for ${\rm C.L.}<0.05$. 

Finally, following the same procedure, we can also
derive a constraint on $\alpha$.  The resulting
C.L. function versus $\alpha$ is given in  Fig.~\ref{fig:deltaalpha} and includes systematic uncertainties.
Ignoring the mirror solution at $\alpha + 180^\circ$, we find
$\alpha \; \in \; (75^\circ, 152^\circ)$ at $68\%$ C.L. 
No constraint on $\alpha$ is achieved at two sigma and beyond.

\section{Conclusions}
\label{sec:conclusions}

We have presented the preliminary measurement of 
\CP-violating asymmetries in $\Btopipipi$ decays dominated by 
the $\rho$ resonance. The results are obtained from a data sample 
of 346 million $\FourS \to B\Bbar$ decays. We perform a time-dependent 
Dalitz plot analysis. From the measurement of the coefficients of 26 form 
factor bilinears we determine the three \CP-violating 
and two \CP-conserving quasi-two-body parameters, where we find a 
$3.0\sigma$ evidence of direct \CP violation. Taking advantage of 
the interference between the $\rho$ resonances in the Dalitz plot,
we derive constraints on the relative strong phase between 
$\Bz$ decays to $\rho^+\pim$ and $\rho^-\pip$, and on the angle
$\alpha$ of the Unitarity Triangle. These measurements are 
consistent with the expectation from the CKM fit~\cite{alphaSM}.

\begin{acknowledgments}
The author wishes to thank the conference organizers for an enjoyable and well-organized workshop.
This work is supported by the  Istituto Nazionale di Fisica Nucleare (INFN) and the 
United State Department of Energy (DOE) under contract DE-AC02-76SF00515.
\end{acknowledgments}

\end{document}